\documentclass[a4paper,11pt]{article}
\usepackage{fullpage}
\usepackage{amsmath,amssymb}
\usepackage{epsfig}
\def\sign{\operatorname{sign}}
\def\tr{\operatorname{tr}}
\def\Tr{\operatorname{Tr}}
\def\Arctan{\operatorname{Arctan}}\def\Arcth{\operatorname{Arcth}}
\def\Li{\operatorname{Li}}
\def\diag{\operatorname{diag}}
\def\negcdot{\negmedspace\cdot\negthinspace}

\begin{document}
\begin{titlepage}
\begin{flushright}

\end{flushright}

\vspace{1cm}
\begin{center}
{\Large \bf  The Rare Decay $D^0 \to \gamma  \gamma  $  }\\

\vspace{1cm}
{\large \bf S. Fajfer$^{a,b}$,  P. Singer$^{c}$, J. Zupan$^a$\\}

\vspace{1cm}
{\it a) J. Stefan Institute, Jamova 39, P. O. Box 3000, 1001 Ljubljana,
Slovenia}
\vspace{.5cm}

{\it b)
Department of Physics, University of Ljubljana, Jadranska 19, 1000
Ljubljana,
Slovenia}
\vspace{.5cm}

{\it c) Department of Physics, Technion - Israel Institute  of
Technology,
Haifa 32000, Israel}

\end{center}

\vspace{0.25cm}

\centerline{\large \bf ABSTRACT}

\vspace{0.2cm}
We present a calculation of the rare decay mode
$ D^0 \to \gamma  \gamma  $, in which the long distance contributions
are
expected to be dominant.
Using the Heavy Quark Chiral  Lagrangian with a
strong $g$
 coupling as
recently determined by CLEO from the $D^* \to D \pi$ width, we
consider both the anomaly contribution which relates to the
annihilation part of the weak Lagrangian
 and  the one-loop $\pi$, $K$ diagrams. The loop contributions which are
 proportional to $g$ and
contain the $a_1$ Wilson coefficient are found to dominate the
decay amplitude, which turns out to be mainly parity violating.
The  branching ratio is then calculated to be
  $(1.0\pm 0.5) \times 10^{-8}$.
Observation of an order of magnitude larger
branching ratio could be indicative of new physics.
\vskip1cm
{ PACS number(s): 13.25.Ft, 14.70.Bh, 12.39.Hg, 12.39.Fe}
\end{titlepage}

\section{Introduction}
  With the new data coming and expected from the B-factories, there is
very
strong emphasis and activity in the field of B-physics in all its
aspects.
This includes the rare decays of B-mesons, which are considered as an
attractive source for possible signals of new physics. In
contradistinction
to the growing efforts to understand the decay mechanisms of rare B
decays,
studies of rare D decays have received less attention in recent years.
Partially this is because theoretical investigations of D weak decays
are
rather difficult, also due to the presence of many resonances close to
this
energy region. The penguin effects on the other hand, which are very
important in B and also in K decays, are  usually suppressed in the case
of charm mesons due to the presence of $d$, $s$, $b$ quarks in the loop
with the
respective values of CKM elements.

  Nevertheless, D meson physics has produced some interesting results in
the past year. Experimental results on time dependent decay rates of
$D^0 \to K^+ \pi^-$ by CLEO \cite{CLEO1} and $D^0 \to K^+ K^-$ and
$D^0 \to K^- \pi^+$ by FOCUS \cite{FOCUS1}
have stimulated several studies on the $D^0 - \bar D^0$ oscillations
\cite{Nir}. The recently measured D* decay width by CLEO \cite{CLEO2}
has provided the long
expected information on the value of $D^*D \pi$ coupling. Among the rare
D
decays, the decays $D\to V
\gamma$ and $D \to V(P) l^+ l^-$  are subjects of CLEO
and FERMILAB searches \cite{E791}. On the theoretical side, these rare
decays
of charm mesons into light vector meson and photon or lepton pair have
been considered lately by several authors (see, e.g.,
\cite{FPS}-\cite{Lebed-00} and
for radiative leptonic D meson decay see \cite{Geng}). The
investigations of
$D\to V \gamma$ showed that certain branching ratios can be as large as
$10^{-5}$, like for $D^0\to \bar{K}^{*0} \gamma$, $D_s^+\to \rho^+
\gamma$ \cite{FPS,Lebed-00}.
However, the decays which are of some relevance to the $D^0\to 2 \gamma$
mode
studied here, like $D^0\to \rho^0 \gamma$, $D^0\to \omega \gamma$, are
expected with
branching ratios in the $10^{-6}$ range \cite{Fajfer-00}. Thus, it is
hard to believe that the
branching ratio of the $D^0\to 2 \gamma$ decay mode can be as high as
$10^{-5}$
in the Standard Model (SM), as found by \cite{RG}. Apart from this
estimation, there
is no other detailed work on $D^0\to 2 \gamma$ in the literature, to the
best of
our knowledge.

 On the other hand, in the B and K meson systems there are numerous
studies
of their decays to two photons. For example, the $B_s\to \gamma \gamma$
decay
has been studied with various approaches within SM and beyond. In SM,
the
short distance (SD) contribution \cite{LSH} leads to a branching ratio
$B(B_s \to \gamma \gamma) \simeq
3.8 \times 10^{-7}$. The QCD corrections enhance this rate to $5 \times
10^{-7}$ \cite{RRS}.
On the other hand, in some of the SM extensions the branching ratio can
be
considerably larger. The two Higgs doublet scenario, for example, could
enhance this branching ratio by an order of magnitude \cite{BI}. Such
"new physics"
effects could at least in principle be dwarfed by long distance (LD)
effects.
However,  existing calculations show that these are not larger than the
SD
contribution \cite{ELLIS}, which is typical of the situation in
radiative B decays \cite{Eilam}. In the $K^0$ system the situation is
rather different. Here, the SD
contribution is too small to account for the observed rates of $K_S\to 2
\gamma$,
$K_L\to 2 \gamma$ by factors of $\sim 3-5$ \cite{Gaillard}, although it
could be of relevance in the
mechanism of CP-violation. Many detailed calculations of these processes
have
been performed over the years (see recent refs.
\cite{Gaillard}-\cite{Kamb} and refs. therein),
especially using the chiral approach to account for the pole diagrams
and
the loops. These LD contributions lead to rates which are compatible
with
existing measurements.

  Motivated by the experimental efforts to observe rare D meson decays,
as
well as by the lack of detailed theoretical treatments, we undertook
an investigation of the $D^0\to  \gamma \gamma$ decay. The short
distance
contribution
is expected to be rather small, as already encountered in the one photon
decays \cite{FPS,BGHP}, hence the main contribution would come from long
distance
interactions. In order to treat the long distance contributions, we use
the
heavy quark effective theory combined with chiral perturbation theory
(HQ$\chi$PT)
\cite{wise}. This approach was used before for treating $D^*$ strong and
electromagnetic
decays \cite{itchpt}-\cite{GS}. The leptonic and semileptonic decays of
D meson were also treated
within the same framework(see \cite{itchpt} and references therein).

   The approach of HQ$\chi$PT introduces several coupling constants that
have
to be determined from experiment. The recent measurement of the $D^*$
decay
width \cite{CLEO2} has determined the $D^*D\pi$ coupling, which is
related to g, the basic
strong coupling of the Lagrangian. There is more ambiguity, however,
concerning
the value of the anomalous electromagnetic coupling, which is
responsible for
the $D^*D\gamma$ decays \cite{stewart,GS}, as we shall discuss later.

   Let us address now some issues concerning the theoretical framework
 used in our treatment. For the weak vertex we shall use the
factorization of
weak currents with nonfactorizable contributions coming from chiral
loops. The
typical energy of intermediate pseudoscalar mesons is of order $m_D/2$,
so that
the chiral expansion $p/\Lambda_\chi$ (for $\Lambda_\chi \gtrsim 1$
GeV)  is rather close to
unity. Thus, for the decay under study here we extend the possible range
of
applicability of the chiral expansion of HQ$\chi$PT, compared to
previous treatments
like $D^*\to D \pi$, $D^*\to D \gamma$ \cite{stewart} or $D^*\to D
\gamma \gamma$ \cite{GS}, in which a heavy meson
appears in the final state, making the use of chiral perturbation theory
rather natural. The suitability of our undertaking here must be
confronted
with experiment, and possibly other theoretical approaches.

  At this point we also remark that the contribution of the order ${\cal
O}(p)$ does
not exist in the $D^0\to \gamma \gamma$ decay, and the amplitude starts
with
contribution of the order ${\cal O}(p^3)$. At this order the amplitude
receives an
annihilation type contribution proportional to the $a_2$ Wilson
coefficient, with the Wess-Zumino anomalous term coupling light
pseudoscalars
to two photons. As we will show, the total amplitude is dominated by
terms
proportional to $a_1$ that contribute only through loops with Goldstone
bosons. Loop contributions proportional to $a_2$ vanish at this order.
We point out that any other model which does not involve intermediate
charged states cannot give this kind of contribution. Therefore, the
chiral
loops naturally include effects of intermediate meson exchange.

  The chiral loops of order ${\cal O}(p^3)$ are finite, as they are in
the similar
case of $K\to \gamma \gamma$ decays \cite{Gaillard}-\cite{Kamb}. The
next to leading terms might be
almost of the same order of magnitude compared to the leading ${\cal
O}(p^3)$ term,
the expected suppression being approximately $p^2/\Lambda^2_\chi$. The
inclusion of next order terms in the chiral expansion is not
straightforward
in the present approach. We include, however, terms which contain the
anomalous
electromagnetic coupling, and appear as next to leading order terms in
the
chiral expansion, in view of their potentially large contribution (as in
$B^*(D^*)\to B(D) \gamma \gamma$ decays considered in \cite{GS}). As it
turns out, these
terms are suppressed compared to the leading loop effects, which at
least
partially justifies the use of HQ$\chi$PT for the decay under
consideration.
Contributions of the same order could arise from light resonances like
$\rho$,
$K^*$, $a_0(980)$, $f_0(975)$. Such resonances are sometimes treated
with hidden
gauge symmetry (see, e.g., \cite{itchpt}), which is not compatible with
chiral perturbation
symmetry. Therefore, a consistent calculation of these terms is beyond
our
scheme and we disregard their possible effect.

 Our paper is organized as follows: in Section  \ref{framework} we
present the basic features
of the model. We give the results and their discussion in Section
\ref{results} and
conclude with a summary in Section \ref{Summary}.

 \section{The theoretical framework}
\label{framework}

The invariant amplitude for $D^0 \to \gamma \gamma $ decay can
 be written using gauge and Lorentz invariance in the following
 form:
\begin{equation}
M = \Big[ i M^{(-)} \big(g^{\mu \nu} -\frac{k_2^\mu k_1^\nu}{k_1
\negcdot k_2} \big)+
 M^{(+)} \epsilon^{\mu \nu\alpha\beta}k_{1\alpha}k_{2\beta}\Big]
 \epsilon_{1\mu}\epsilon_{2\nu},\label{eq-104}
\end{equation}
where $M^{(-)}$ is a parity violating  and $M^{(+)}$
a parity conserving  part of the amplitude,
while $k_{1(2)}$, $\epsilon_{1(2)}$ are respectively the
four momenta and the polarization vectors of the outgoing
photons.

In the  discussion of weak radiative decays
$q' \to q \gamma \gamma$ or $q' \to q \gamma $ decays,
usually
the short (SD) and long distance (LD) contribution are separated.
The SD contribution in these transitions is a  result of the
penguin-like transition, while the long distance contribution
arises in
particular pseudoscalar meson decay as a result of the
nonleptonic four quark weak Lagrangian, when the photon is emitted
from the
quark legs. Here we follow this classification.
In the case of $b \to s \gamma \gamma$ decay
\cite{HK}  it was noticed that
without QCD corrections the rate
$\Gamma (b \to s \gamma \gamma)/ \Gamma (b \to s \gamma)$
is about $10^{-3}$. One expect that  a similar effect will show up
in the case of $c \to u \gamma \gamma$ decays.
Namely, according to the result of \cite{HK}
the largest contribution to $c \to u \gamma \gamma$ amplitude
would arise from the  photon emitted either from $c$ or $u$
quark legs
in the case of the  penguin-like  transition
$c \to u \gamma$.
Without QCD corrections the branching ratio for
$c \to u \gamma$  is rather suppressed,
being of the order $10^{-17}$ \cite{BGHP,HP}.
The QCD corrections \cite{GMW} enhance it up to order of $10^{-8}$.

 In our approach we include the $c \to u \gamma$ short distance
contribution by using the Lagrangian
 \begin{equation}
 {\cal L}=-\frac{G_f}{\sqrt{2}}V_{us} V_{cs}^*
C_{7\gamma}^{eff}\frac{e}{4\pi^2}
F_{\mu \nu}
  m_c \; \big(\bar u \sigma^{\mu\nu} \tfrac{1}{2}(1 + \gamma_5) c\big),
\label{eq-103}
 \end{equation}
 where $m_c$ is a charm quark mass.  In our analysis we follow
 \cite{GMW, Sasathesis}
 and we take $C_{7\gamma}^{eff}= (-0.7+ 2 i) \times 10^{-2}$.

 The main LD contribution will arise from the   effective four quark
nonleptonic $\Delta C = 1$
  weak Lagrangian
given by
\begin{equation}
{\cal L}=-\frac{G_f}{\sqrt{2}} \sum_{q=d,s}V_{uq}
 V_{cq}^* \big[ a_1  \big(\bar{q} \Gamma^\mu c) (\bar{u}
\Gamma_\mu q)+
 a_2 (\bar{u}\Gamma^\mu c) (\bar{q} \Gamma_\mu q)\big],
\label{eq-107}
\end{equation}
where $\Gamma^\mu=\gamma^\mu(1-\gamma_5)$, $a_i$ are effective
Wilson
coefficients \cite{buras}, and $V_{q_i q_j}$ are CKM matrix
elements. At
this point it is worth pointing out that long distance interactions will
contribute only if the $SU(3)$
flavor symmetry  is broken, i.e. if $m_s\neq m_d$. Namely, due to
$V_{ud}V_{cd}^* \simeq -V_{us}V_{cs}^*$, if  $m_d = m_s$
the contributions arising from the weak Lagrangian (\ref{eq-107})
cancel. 

Now, we turn to describe  some of the basic features of the HQ$\chi$PT.
This model would serve us  as a
hadronized counterpart of the quark effective weak
Lagrangian. One has
the  usual ${\cal O}(p^2)$ chiral Lagrangian for the light
pseudoscalar mesons
\begin{equation}
{\cal L}_{\text str}^{(2)}=\frac{f^2}{8} \tr (\partial^\mu \Sigma
\partial_\mu \Sigma^\dagger)+\frac{f^2 B_0}{4}\tr({\cal M}_q \Sigma
+{\cal M}_q\Sigma^\dagger) \; ,\label{eq-11}
\end{equation}
where $\Sigma = \exp{(2 i \Pi/f)}$ with
$\Pi =  \sum_j \frac{1}{\sqrt{2}}\lambda^j \pi^j$ containing the
Goldstone
bosons $\pi,
K,
\eta$, while the trace $\tr$ runs over flavor indices and ${\cal
M}_q=\diag
(m_u,m_d,m_s)$ is the current quark mass matrix. From this Lagrangian,
 we can deduce the
 light weak current of the order ${\cal O}(p)$
\begin{equation}
j_\mu^X \, = \, -i\frac{f^2}{4}\tr(\Sigma \partial_\mu
\Sigma^\dagger\lambda^X) \; ,
\label{jX}
\end{equation}
corresponding to the quark current
$j_\mu^X=\bar{q}_{L}\gamma_\mu\lambda^Xq_{L}$. ($\lambda^X$ is an SU(3)
flavor matrix.)

In the heavy meson sector interacting with light mesons we have
the following lowest order ${\cal O}(p)$  chiral Lagrangian
\begin{equation}
{\cal L}_{\text str}^{(1)}=-\Tr(\bar{H}_{a}iv\negcdot D_{ab}H_{b})+g
\Tr(\bar{H}_{a}H_{b} \gamma_\mu {\cal A}_{ba}^\mu \, \gamma_5) \; ,
\label{eq-8}
\end{equation}
where $ D_{ab}^\mu H_b=\partial^\mu H_a - H_b{\cal V}_{ba}^\mu$,
 while the trace $\Tr$ runs over Dirac indices.
Note that in (\ref{eq-8}) and the rest of this section
$a$ and $b$ are {\em flavor} indices.

The vector and axial vector fields
${\cal V}_{\mu}$ and
${\cal A}_\mu$ in (\ref{eq-8}) are given by:
\begin{equation}
{\cal V}_{\mu} = \frac{1}{2}(\xi\partial_\mu\xi^\dagger
+\xi^\dagger\partial_\mu\xi) \qquad   \qquad
{\cal A}_\mu = \frac{i}{2}
(\xi^\dagger \partial_\mu\xi -\xi\partial_\mu\xi^\dagger) \; ,
\label{defVA}
\end{equation}
where $\xi = \exp{(i \Pi/f)}$. The heavy meson field
 $H_{a}$ contains
 a spin zero and spin one boson
\begin{align}
H_{a} =& P_+ (P_{\mu a} \gamma^\mu -
 P_{5 a} \gamma_5),\\
\overline{H}_{a}=&\gamma^0 (H_{a})^\dagger \gamma^0
=\left[ P_{\mu a}^{\dagger} \gamma^\mu
+ P_{5 a}^\dagger \gamma_5\right] P_+ \; ,
\label{barH}
\end{align}
with $ P_{\pm}=(1 \pm \gamma^\mu v_\mu)/2 $ being the  projection
operators.
The field $P_5(P_5^\dagger)$ annihilates (creates) a pseudoscalar
meson  with
a heavy quark  having  velocity $v$,
and similar for spin one mesons.

For a decaying heavy quark, the
 weak current is given by
\begin{equation}
J_a^\lambda =\overline{q}_a \gamma^\lambda L Q \; ,
\label{Lcur}
\end{equation}
where $L=(1-\gamma_5)/2$ and $Q$ is the heavy quark field in the
 full theory, in our case a $c$-quark, and $q$ is the light quark
 field.

From symmetry grounds,
 the heavy-light weak current is  bosonized
in the following way \cite{wise}
\begin{equation}
J^\lambda_a =  \frac{i\alpha}{2} \Tr [\gamma^\lambda \, L \, H_{b}\,
\xi_{ba}^\dagger]
\; ,
\label{JH}
\end{equation}
where $\alpha$ is related to the
 physical decay constant $f_D$ through the well known matrix
 element
\begin{equation}
\langle  0| \overline{u}\gamma^\lambda \gamma_5 c|D^0 \rangle
= -2 \langle  0| J_a^\lambda|D^0 \rangle
=im_D v^\lambda f_D \; .
\label{fD}
\end{equation}
Note that the current (\ref{JH}) is  ${\cal O}(p^0)$ in the chiral
counting.

In the  calculation of short distance contribution \eqref{eq-103}
there
appears  the operator $ (\bar{u} \sigma_{\mu\nu}
\tfrac{1}{2} (1+\gamma_5)c)$.
Using heavy quark symmetry this operator can be translated into
an operator containing meson fields only \cite{Grinstein-94}
\begin{equation}
\big(\bar{u} \sigma_{\mu\nu} \tfrac{1}{2}(1+
\gamma_5)c\big)\to \frac{i
\alpha}{2}\Tr[\sigma_{\mu\nu}\tfrac{1}{2}
(1+\gamma_5)H_b\xi_{ba}^\dagger].\label{eq-105}
\end{equation}

The photon couplings  are obtained by gauging the Lagrangians
\eqref{eq-11}, \eqref{eq-8} and the light current \eqref{jX}
with the $U(1)$ photon field $B_\mu$.
The covariant derivatives are
then ${\cal D}_{ab}^\mu H_b=
\partial^\mu H_a +i e B^\mu (Q'H-H Q)_a-H_b {\cal V}_{ba}^\mu$
and
${\cal D}_\mu=\partial_\mu \xi +
i e B_\mu [Q,\xi]$ with $Q=\diag (\frac{2}{3},-\frac{1}{3},
-\frac{1}{3})$
and $Q'=\frac{2}{3}$ (for our case of $c$ quark). The vector and
axial vector
fields  \eqref{defVA} change after gauging and now they read $ {\cal
V}_{\mu}
=\frac{1}{2}(\xi{\cal D}_\mu\xi^\dagger
+\xi^\dagger{\cal D}_\mu\xi) $ and
$ {\cal A}_\mu = \frac{i}{2}
(\xi^\dagger {\cal D}_\mu\xi -\xi{\cal D}_\mu\xi^\dagger)$.
The light weak current \eqref{jX} contains after gauging the covariant
derivative ${\cal D}_\mu$ instead
of $\partial_\mu$. However, the gauging procedure alone does not
introduce
a coupling between heavy vector and
pseudoscalar meson fields and the photon without emission or absorption
of
additional Goldstone boson,
which is needed to account, for example, for $D^*\to D \gamma$. To
describe
this
electromagnetic interaction
we follow \cite{stewart} introducing an additional gauge invariant
contact
term with
an unknown coupling $\beta$
of dimension -1.
\begin{equation}
{\cal  L}_\beta=-\frac{\beta e}{4} \Tr \bar{H}_a H_b \sigma^{\mu
\nu}F_{\mu
\nu} Q_{ba}^\xi -\frac{e}{4 m_Q} Q' \Tr \bar{H}_a
\sigma^{\mu \nu} H_a
F_{\mu\nu},\label{eq-100}
\end{equation}
where $Q^\xi=\frac{1}{2}(\xi^\dagger Q \xi+\xi Q \xi^\dagger)$ and
$F_{\mu\nu}=\partial_\mu B_\nu- \partial_\nu B_\mu$.
The first term concerns the contribution
of the light quarks in
the heavy meson and the  second term
describes  emission of
a photon
from the heavy quark. Its coefficient  is fixed by heavy quark
symmetry. From this "anomalous" interaction, both $H^* H \gamma$ and
$H^*H^*\gamma$ interaction terms arise. Even though the
Lagrangian \eqref{eq-100} is formally $1/m_Q \sim m_q$ suppressed,
we
do not neglect it, as  it has been found that it gives a sizable
contribution to $D^*(B^*)\to D(B) \gamma\gamma$ decays \cite{GS}. In the
case of $D^0\to \gamma\gamma$ it gives largest contribution to the
parity conserving part
of the
amplitude, however it does not contribute to the decay rate
 by more than $10\%$, as
 will be shown later. The Lagrangian \eqref{eq-100}
in principle
receives a number of other contributions at the order of $1/m_Q$,
however, these
can be absorbed in the definition of $\beta$ for the processes
considered
\cite{stewart}.

\section{Results}
\label{results}

\begin{figure}
\begin{center}
\epsfig{file=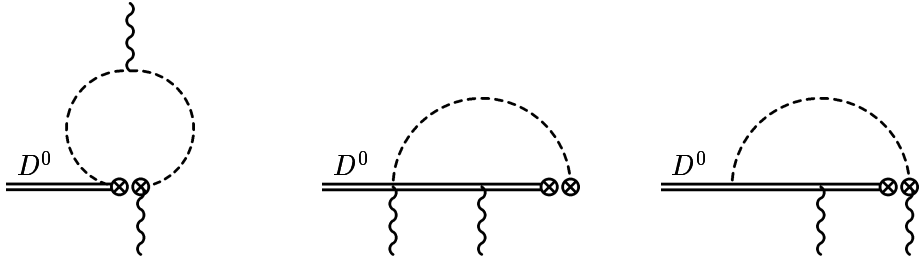}
\caption{One loop diagrams (not containing $\beta$-like  terms
\eqref{eq-100}) that give vanishing contributions. The dashed line
represents
charged
Goldstone bosons flowing in the loop ($K^+,\pi^+$), while the
double line represents  heavy mesons, $D$ and $D^*$. }\label{fig-1}
\end{center}
\end{figure}

\begin{figure}
\begin{center}
\epsfig{file=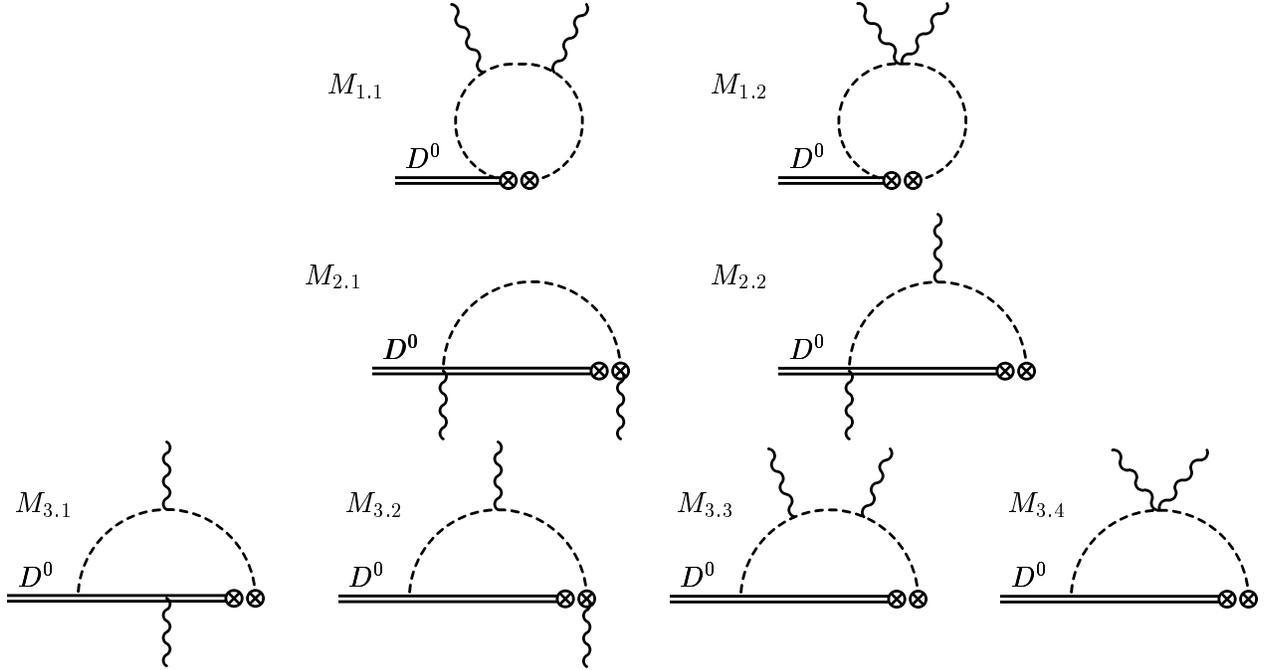}
\caption{One loop diagrams, not containing beta-like terms
\eqref{eq-100},  that give nonvanishing contributions to the
$D^0\to
\gamma \gamma$ decay amplitude. Each sum of the  amplitudes on
diagrams in one row $M_i=\sum_j
M_{i.j}$ is gauge invariant and finite. Numerical values are
listed in
Table \ref{tab-1}. }\label{fig-2}
\end{center}
\end{figure}

\begin{figure}
\begin{center}
\epsfig{file=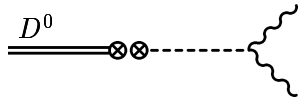}
\caption{Anomalous contributions to $D^0\to \gamma\gamma$ decay.
The intermediate pseudoscalar mesons propagating from the weak
vertex are $\pi^0,
\eta,\eta'$. }\label{fig-3}
\end{center}
\end{figure}

 The decay width for the $D^0\to \gamma \gamma$ decay
can be obtained using the amplitude decomposition in \eqref{eq-104}:
\begin{equation}
\Gamma_{D^0\to \gamma \gamma}= \frac{1}{16 \pi m_D} ( |M^{(-)}|^2+
\frac{1}{4} |M^{(+)}|^2 m_D^4) . \label{eq-106}
\end{equation}
The short distance contribution to $D^0\to \gamma\gamma$ decay
width is
estimated using $c\to u \gamma$ effective Lagrangian
\eqref{eq-103},
\eqref{eq-105} with one photon emitted from $D^0$ leg via ${\cal
L}_\beta$ term \eqref{eq-100}. The parity violating part of the
short distance amplitude is
\begin{equation}
   M_{SD}^{(-)}= \frac{m_D^{3/2}}{12 \pi^2} \frac{G_f}{\sqrt{2}}
V_{us} V_{cs}^* C_{7\gamma}^{eff} e^2
(\beta m_c+1)\alpha \frac{1}{1+2 \Delta^*/m_D},
\end{equation}
while the parity conserving  part of the amplitude is
\begin{equation}
 M_{SD}^{(+)}=\frac{\sqrt{m_D}}{12 \pi^2} \frac{G_f}{\sqrt{2}}
V_{us} V_{cs}^* C_{7\gamma}^{eff}
e^2
(\beta m_c+1) \alpha \frac{2}{m_D+2 \Delta^*}, \label{eq-3}
\end{equation}
where $\Delta^{*}=m_{D^{0*}}-m_{D^0}$.
Turning now to the long distance contributions, we depict in Figs.
\ref{fig-1} and \ref{fig-2} the loop diagrams
arising to leading order ${\cal O }(p^3)$ by using
Eqs.\eqref{eq-11}-\eqref{eq-8} and \eqref{JH}. The circled
cross indicates the weak
interaction. In Figure \ref{fig-1} we grouped all diagrams which vanish
by
 symmetry considerations. All
nonvanishing contributions are assembled in Fig. \ref{fig-2}. We denote
the gauge invariant sums  corresponding
to nonvanishing diagrams of Fig. \ref{fig-2} by $M_i^{(\pm)}=\sum_j
M_{i.j}^{(\pm)}$ (the gauge invariant sums are sums of diagrams in each
row of Fig. \ref{fig-2}), where $+ (-)$ denotes parity conserving
(violating) part of the amplitude, as in \eqref{eq-106}.  The parity
violating  sums, which
arise from the first term in \eqref{eq-107}  are \\
\begin{align}
M_1^{(-)}&= -\frac{(m_D)^{3/2}}{4 \pi^2} \frac{G_f}{\sqrt{2}}
a_1
\alpha e^2 \Big[V_{us} V_{cs}^*
M_4(m_K,-\tfrac{m_D^2}{2})+V_{ud}V_{cd}^*
M_4(m_\pi, -\tfrac{m_D^2}{2})\Big]\label{eq-1},\\
\begin{split}
M_2^{(-)}&=  \sqrt{m_D} \frac{G_f}{\sqrt{2}} a_1 e^2 g \alpha
\frac{1}{8 \pi^2} \Big[\\
&\qquad V_{us} V_{cs}^*( \frac{1}{m_D/2+\Delta_s^* }
I_2(m_K,\tfrac{m_D}{2}+\Delta_s^*) -
2G_3(m_K,m_D+\Delta_s^*,-\tfrac{m_D}{2}))+\\
&\qquad V_{ud} V_{cd}^*( \frac{1}{m_D/2+\Delta_d^* }
I_2(m_\pi,\tfrac{m_D}{2}+\Delta_d^*) -
2G_3(m_\pi,m_D+\Delta_d^*,-\tfrac{m_D}{2}))\Big],
\end{split}\\
M_3^{(-)}&=  \sqrt{m_D}  \frac{G_f}{\sqrt{2}} a_1 g e^2 \alpha
\frac{1}{2 \pi^2} \Big[
V_{us}V_{cs}^* f(m_K,\Delta_s^*,m_D) +V_{ud} V_{cd}^*
f(m_\pi,\Delta_d^*,m_D)\Big], \label{eq-2}
\end{align}
with
\begin{equation}
\begin{split}
f(m,&\Delta,m_D)= \frac{m^2}{m_D} \Big[
G_0\big(m,\Delta+\tfrac{m_D}{2},\tfrac{m_D}{2}\big)-\frac{1}{2}
G_0(m,\Delta,\tfrac{m_D}{2})\Big]+\frac{5 m_D}{8} + \frac{\Delta}{2}\\
& +\frac{(m^2-\Delta^2)}{2} \Big[\frac{1}{m_D} N_0(m,m_D^2)
+\Big(\frac{1}{2}+\frac{\Delta}{m_D}\Big)\; \overline{G}_0\big(m,
\Delta+\tfrac{m_D}{2},m_D\big)+m^2\;
\overline{M}_0\big(m,\Delta+\tfrac{m_D}{2},m_D\big)\Big]\\
&+ \Big(\Delta-\frac{m_D}{2}\Big)
M_2(m,-\tfrac{m_D^2}{2})-\frac{1}{4}\Big(\frac{m_D}{2}-\Delta\Big)
N_0(m,m_D^2)\\
&+ \frac{(2\Delta+m_D)}{4 m_D (\Delta+m_D)}
I_2(m,\Delta+m_D)-\frac{(3m_D^2/2+3
\Delta m_D +2 \Delta^2-2 m^2)}{2 m_D^2 (m_D+2\Delta)}
I_2(m,\Delta+\tfrac{m_D}{2})\\
&+ \frac{(m_D \Delta -2 m^2 +
2 \Delta^2)}{4 m_D^2 \Delta} I_2(m,\Delta).
\end{split}
\end{equation}
The parity conserving parts of the amplitude $M_i^{(+)}$ vanish
for the diagrams on Fig. \ref{fig-2}.
We denoted  $\Delta_q^{(*)}=m_{D_q^{(*)}}-m_{D^0}$,
while functions
$I_2(m,\Delta)$, $G_0(m,\Delta,v\negcdot k)$,
$G_3(m,\Delta,v\negcdot k)$, $N_0(m,k^2)$,
$\overline{G}_0(m,\Delta,v\negcdot k)$,
$\overline{M}_0(m,\Delta,v \negcdot k)$, $M_2(m,k_1\negcdot k_2)$,
 $M_4(m,k_1 \negcdot k_2)$ are presented in the Appendix \ref{app-A}.
 Note that the sums of amplitudes \eqref{eq-1}-\eqref{eq-2}
 are  gauge invariant and finite.
This is  expected, since one cannot generate counterterms at
 this order. There is  no $\mu$ dependence apart
 from the one hidden in $a_1$ even though $\mu$ appears in
 the above functions, but it
 cancels out completely. Note also, that the one
 loop chiral corrections vanish in the exact $SU(3)$ limit,
 i.e. when $m_K \to m_\pi$, as it is expected.
One should note that taking the chiral limit (i.e. $m_s, m_d \to 0$)
is not unambiguous. Namely, in the combined heavy quark effective theory
and
the chiral perturbation theory, beside chiral logarithms
there are also functions of
the form $F(m_q/\Delta)$ whose value depend on the way one takes the
limit (see e.g. Ref. \cite{BG}).

We remark that there exist additional diagrams of the same order in
the chiral
expansion as ones  given on Fig. \ref{fig-2}, but proportional to
 $a_2$ part of the effective
weak Lagrangian \eqref{eq-107}. In these additional diagrams the chiral
loop is attached to  the light current in the factorized vertex, while
the  photons are
 emitted from the
pseudoscalars in the loop, or they come from the weak vertex.
However,   the  amplitudes of these diagrams
 vanish due to Lorentz symmetry.

The contribution coming from the anomalous coupling
$\pi^0 \gamma \gamma$, $\eta\gamma \gamma$, $\eta'\gamma \gamma$
(Fig. \ref{fig-3}) is
\begin{equation}
\begin{split}
M_{Anom.}^{(+)}=-&\sqrt{m_D} \frac{G_f}{\sqrt{2}}a_2 \alpha
\frac{e^2}{4 \pi^2}
\sum_{P=\pi^0,\eta,\eta'} \frac{m_D}{m_D^2-m_P^2} K_P\\
&K_{\pi^0}=V_{ud}V_{cd}^*\\
& K_\eta=\big[V_{ud}V_{cd}^*\big(\tfrac{\sin
\Theta}{\sqrt{3}}-\tfrac{\cos
\Theta}{\sqrt{6}}\big)+V_{us}V_{cs}^*\big(\tfrac{\sin
\Theta}{\sqrt{3}}+\tfrac{\sqrt{2}\cos
\Theta}{\sqrt{3}}\big)\big]\big[\tfrac{\sqrt{2}\cos
\Theta}{\sqrt{3}}-\tfrac{4\sin
\Theta}{\sqrt{3}}\big]\\
& K_\eta'=\big[-V_{ud}V_{cd}^*\big(\tfrac{\sin
\Theta}{\sqrt{6}}+\tfrac{\cos
\Theta}{\sqrt{3}}\big)+V_{us}V_{cs}^*\big(\tfrac{\sqrt{2}\sin
\Theta}{\sqrt{3}}
-\tfrac{\cos
\Theta}{\sqrt{3}}\big)\big]\big[\tfrac{\sqrt{2}\sin\Theta}{\sqrt{3}}
+\tfrac{4\cos \Theta}{\sqrt{3}}\big],
\end{split}
\end{equation}
where  $\theta=-20^o \pm 5^o $ is the $\eta -\eta'$ mixing angle
and we have set $f_\pi=f_{\eta_8}=f_{\eta_0}$. This choice of
parameters reproduces the experimental results for the
$\pi^0\to \gamma \gamma$, $\eta \to \gamma \gamma$, and
$\eta'\to \gamma \gamma$ decay width \cite{PDG-00}.
In the  numerical evaluation  we use the values of $\alpha$ and  $g$
obtained
 within the same framework as in
 \cite{itchpt,stewart,Grinstein-92,Grinstein-94,BG}.
The coupling $g$ is extracted from existing
experimental data  on $D^*\to D\pi$.  Recently CLEO Collaboration has
obtained the
first measurement of $D^{*+}$ decay width $\Gamma(D^{*+})=96\pm 4\pm 22$
$\rm{ keV}$ \cite{CLEO2} by studying the $D^*\to D^0 \pi^+$. Using the
value of
decay width together with
branching ratio $Br(D^{*+}\to D^0 \pi^+)=(67.7\pm 0.5)\% $   one
immediately finds at tree level that $g=0.59 \pm 0.08 $.
The chiral corrections to this coupling were found to contribute about
$10\%$  \cite{itchpt,stewart}.  In order
to obtain the $\alpha$ coupling, we use present experimental
data on
$D_s$ leptonic decays.  Namely, at the tree level there is
a relation
$f_{D}=f_{D_{s}}=\alpha/\sqrt{m_D}$. From the experimental branching
ratio
$D_s\to\mu \nu_\mu$ and the  $D_s$ decay width \cite{PDG-00} one gets
$f_{D_s}=0.23\pm 0.05$ $ {\rm GeV}$ and $\alpha=0.31\pm 0.04 \; {\rm
GeV}^{3/2}$. The $SU(3)$ breaking effects in the form of chiral loops
and the counterterms can change the extracted value of $\alpha$.
One chiral loop corrections can amount to about
$40\%$ when   $g$ is taken to be $0.59$.
This value might be  changed
by the finite part of the counterterms.
However, the contributions coming from
counterterms are not known and due to the lack of experimental data
they cannot be fixed yet.  In our
calculation we take $\alpha=0.31\; {\rm GeV}^{3/2}$, keeping in
mind that
the chiral corrections might be important (for instance  setting
counterterms to zero in one loop calculation one gets $\alpha=0.21\pm
0.04$ $ {\rm
GeV}^{3/2}$  using $g=0.59$, for details see
Appendix B of \cite{Eeg:2001un}).
For the Wilson coefficients $a_1$ we take $1.26$
and $a_2=-0.47$ \cite{buras}.
We present the  numerical results
 for the  one  loop amplitudes in Table \ref{tab-1}.
\begin{table} [h]
\begin{center}
\begin{tabular}{|l|r l|r l|} \hline
 &$M_{ i}^{(-)}$&$ [\times 10 ^{-10}{\rm  \;GeV}]$& $M_{i}^{(+)}$&$
[\times10
^{-10}{\rm  \;GeV^{-1}}]$\\ \hline\hline
Anom. & $0$ & & $-0.53$& \\ \hline
SD & $-0.27$&$-0.81 i$ & $-0.16$&$ -0.47 i$ \\ \hline
$1$ & $3.55$&$+9.36i$ & $0$&\\ \hline
$2$ &$1.67$ & &  $0$&  \\ \hline
$3$ & $-0.54$&$+2.84i$&$0$& \\ \hline
\hline
$\sum_i  M_i^{(\pm)}$& $4.41$&$+11.39 i$ &$-0.69$&$ -0.47 i$\\ \hline

\end{tabular}
\caption{\footnotesize{Table of the nonvanishing finite
amplitudes. The amplitudes coming from the
anomalous and short distance
($C_{7\gamma}^{eff}$) Lagrangians are presented. The finite
and  gauge invariant sums of
one-loop amplitudes are listed in the next three lines
($M_i^{(\pm)}=\sum_j M_{i.j}^{(\pm)}$).
The numbers $1,2,3$ denote the row of diagrams on the
Fig. \ref{fig-2}.
In the last line the sum of all amplitudes is given.}}
\label{tab-1}
\end{center}
\end{table}

\begin{figure}
\begin{center}
\epsfig{file=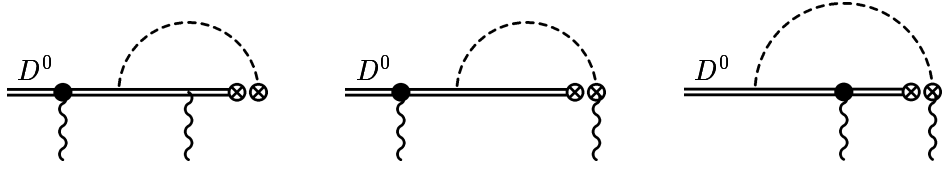}
\caption{The diagrams with one $\beta$-like \eqref{eq-100} coupling
(described by $\bullet$), which give
vanishing amplitudes. }\label{fig-4}
\end{center}
\end{figure}

\begin{figure}
\begin{center}
\epsfig{file=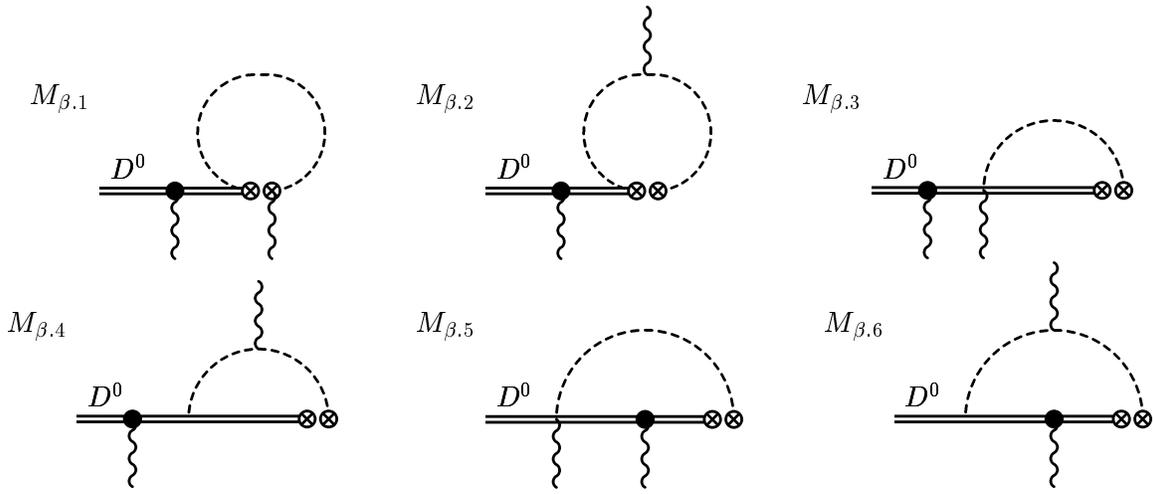}
\caption{The diagrams which give nonzero amplitudes
with one $\beta$-like coupling. }\label{fig-5}
\end{center}
\end{figure}

In the determination of $D^*\to D \gamma\gamma $ and
$B^* \to B \gamma \gamma$ a sizable contribution from
$\beta$-like electromagnetic terms \eqref{eq-100} has been
found \cite{GS}. Therefore we have to investigate their
effect
in the $D^0 \to \gamma \gamma $ decay amplitude. The considerations of
Eq. \eqref{eq-100} gives us additional diagrams which are given in Fig.
\ref{fig-4} and \ref{fig-5}, where the $\beta$ vertex is indicated by
$\bullet$.
The nonzero parity violating  parts of the one loop
diagrams containing
$\beta$
coupling are (Fig. \ref{fig-5})
\begin{align}
\begin{split}
M_{\beta .4}^{(-)}&= \sqrt{m_D}\frac{G_f}{\sqrt{2}}a_1 e^2 g
\alpha
(\beta
+\frac{1}{m_c}) \frac{1}{(m_D +2 \Delta^*)} \frac{1}{16 \pi^2}
\frac{m_D^2}{3}\\
&\qquad \qquad\Big[ V_{us}V_{cs}^*
G_3(m_K,m_D+\Delta_s^*,-\tfrac{m_D}{2})+
V_{ud}V_{cd}^*G_3(m_\pi,m_D+\Delta_d^*,-\tfrac{m_D}{2})\Big],
\end{split}\\
\begin{split}
M_{\beta .6}^{(-)}&= \frac{G_f}{\sqrt{2}} a_1 e^2 g \alpha
(\beta-\frac{2}{m_c})
\frac{(m_D)^{\frac{3}{2}}}{48 \pi^2}\Big\{ V_{us}V_{cs}^*
\Big[G_3(m_K,\tfrac{m_D}{2}+\Delta_s^*,\tfrac{m_D}{2})-G_3(m_K,\Delta_s^*,\tfrac{m_D}{2})\Big]\\
&\qquad \qquad \qquad \qquad + V_{ud}V_{cd}^*
\Big[G_3(m_\pi,\tfrac{m_D}{2}+\Delta_d^*,
\tfrac{m_D}{2})-G_3(m_\pi,\Delta_d^*,\tfrac{m_D}{2})\Big]\Big\},
\end{split}
\end{align}
while the parity conserving parts of the amplitudes arising from the
one loop diagrams with $\beta$
coupling are
\begin{align}
M_{\beta .1}^{(+)}&=\frac{G_f}{\sqrt{2 m_D}}a_1 e^2 \alpha
(\beta+\frac{1}{m_c})
\frac{1}{m_D+2 \Delta^*} \frac{1}{12 \pi^2}\Big[ V_{us}V_{cs}^*
I_1(m_K)+V_{ud}V_{cd}^* I_1(m_\pi)\Big],\\
M_{\beta .2}^{(+)}&=-\frac{G_f}{\sqrt{2 m_D}}a_1 e^2 \alpha
(\beta+\frac{1}{m_c})\frac{1}{(m_D+2\Delta^*)}\frac{1}{12\pi^2}
\Big[V_{us}V_{cs}^* I_1(m_K)+V_{ud}V_{cd}^*
I_1(m_\pi)\Big]=-M_{\beta.1}^{(+)},\\
\begin{split}
M_{\beta .3}^{(+)}&=-\frac{G_f}{\sqrt{2 m_D}} a_1 e^2 g
\alpha(\beta+\frac{1}{m_c})
\frac{1}{m_D+ 2 \Delta^*} \frac{1}{12\pi^2}\Big\{ V_{us}V_{cs}^*\big[
I_2(m_K,m_D+\Delta_s) +I_1(m_K)\big]+\\
&\qquad \qquad \qquad \qquad
+V_{ud}V_{cd}^*\big[I_2(m_\pi,m_D+\Delta_d)+I_1(m_\pi)\big]\Big\},
\end{split}\\
\begin{split}
M_{\beta .4}^{(+)}&=\frac{G_f}{\sqrt{2 m_D}}a_1 e^2 g \alpha
(\beta
+\frac{1}{m_c})
\frac{1}{m_D+2 \Delta^*} \frac{1}{6\pi^2}\Big\{\\
&\qquad\qquad V_{us}V_{cs}^* \big[\frac{1}{2}
I_1(m_K)+(m_D+\Delta_s)G_3(m_K,m_D+\Delta_s,-\tfrac{m_D}{2})\big]+\\
&\qquad \qquad + V_{ud}V_{cd}^* \big[\frac{1}{2}
I_1(m_\pi)+(m_D+\Delta_d)G_3(m_\pi,m_D+
\Delta_d,-\tfrac{m_D}{2})\big]\Big\},
\end{split}
\end{align}
\begin{align}
\begin{split}
M_{\beta .5}^{(+)}&=\frac{G_f}{\sqrt{2m_D}} a_1 e^2
g \alpha(-\beta
+\frac{2}{m_c})
\frac{1}{24 \pi^2}\Big\{\\
&\qquad \qquad V_{us}V_{cs}^* \frac{1}{m_D+2
(\Delta_s-\Delta_s^*)}\big[I_2(m_K,m_D+\Delta_s)-I_2(m_K,\tfrac{m_D}{2}+\Delta_s^*)\big]+\\
&\qquad \qquad + V_{ud}V_{cd}^* \frac{1}{m_D+2
(\Delta_d-\Delta_d^*)}\big[I_2(m_\pi,m_D+\Delta_d)-
I_2(m_\pi,\tfrac{m_D}{2}+\Delta_d^*)\big]\Big\},
\end{split}\\
\begin{split}
M_{\beta .6}^{(+)}&=\frac{G_f}{\sqrt{2 m_D}}a_1 e^2 g \alpha
(\beta-\frac{2}{m_c})
\frac{1}{24 \pi^2}\Big\{\\
V_{us}V_{cs}^* & \Big[-\frac{(\frac{m_D}{2}+
\Delta_s^*)}{(\frac{m_D}{2}+\Delta_s-\Delta_s^*)}G_3\big(m_K,
\tfrac{m_D}{2}+\Delta_s^*,-\tfrac{m_D}{2}\big)+
\frac{(m_D+\Delta_s)}{(\frac{m_D}{2}+\Delta_s-
\Delta_s^*)}G_3\big(m_K,m_D+\Delta_s,-\tfrac{m_D}{2}\big)\Big]+\\
+V_{ud}V_{cd}^*& \Big[-\frac{(\frac{m_D}{2}+
\Delta_d^*)}{(\frac{m_D}{2}+\Delta_d-
\Delta_d^*)}G_3\big(m_\pi,\tfrac{m_D}{2}+\Delta_d^*,-
\tfrac{m_D}{2}\big)+\frac{(m_D+\Delta_d)}{(\frac{m_D}{2}+
\Delta_d-\Delta_d^*)}G_3\big(m_\pi,m_D+\Delta_d,-
\tfrac{m_D}{2}\big)\Big]\Big\}.
\end{split}
\end{align}
The amplitudes with  $\beta$ coupling are not finite and have to be
regularized. We use the strict $\overline{MS}$ prescription
 $\bar{\Delta}=1$ as in \cite{stewart} (note that in \cite{itchpt}
$\bar{\Delta}=0$ has been  used). We  take  $\mu = 1$ $\rm{GeV}
\simeq
\Lambda_\chi$ as in \cite{stewart}.

  In order
to obtain the value of $\beta$ we use the available experimental
data from $D^{*+}\to D^+ \gamma$, $D^{*0}\to D^{0}\gamma$ decays.
For instance, one can  use the recently determined $D^{*+}$ decay
width  $\Gamma (D^{*+})=96\pm 4\pm 22\; {\rm keV}$ \cite{CLEO}
together with the branching ratio
$Br(D^{*+}\to D^+ \gamma)=(1.6 \pm 0.4)\%$ \cite{PDG-00}.
At tree level one has
\begin{equation}
\Gamma(D^{*+}\to D^+ \gamma)=\frac{e^2}{12 \pi} \big(
\tfrac{2}{3} \tfrac{1}{m_c}-\tfrac{1}{3}
\beta\big)^2 k_\gamma^3 ,\label{eq-102}
\end{equation}
with $k_\gamma=\frac{m_{D^*}}{2}(1-\frac{m_D^2}{m_{D^*}^2})$ the
momentum of
outgoing photon. Using the experimental data and $m_c=1.4 \;
{\rm GeV}$
one
arrives at $\beta=2.9\pm 0.4 \; {\rm GeV}^{-1}$
\renewcommand{\thefootnote}{\fnsymbol{footnote}}
\footnote{There is also
a solution  of \eqref{eq-102} $\beta=0.09 \pm 04 \;
{\rm GeV}^{-1}$ that,
however,
does not agree with the determination of $\beta$
from $D^{*0}$ decay.},
where
the errors reflect the experimental errors.

On the other hand one can also use the ratio of partial decay width in
$D^{*0}$
system $\Gamma(D^{*0}\to D^0\gamma):\Gamma(D^{*0}\to D^0 \pi^0)=(38.1\pm
2.9):(61.9\pm 2.9)$, where the experimental errors are considerably
smaller
than in the previous case.  At tree level one has
\begin{equation}
\frac{ \Gamma(D^{*0}\to D^0\gamma)}{\Gamma(D^{*0}\to D^0
\pi^0)}=\frac{e^2}{12 \pi} \frac{ k_\gamma^3}{k_\pi^3}\frac{12 \pi
f^2}{g^2}
\big(\tfrac{2}{3}\beta +\tfrac{2}{3} \tfrac{1}{m_c}\big)^2 ,
\label{eq-108}
\end{equation}
with $k_\gamma$,$k_\pi$ the momenta of outgoing photon and pion
respectively.
Using $m_c=1.4 \; {\rm GeV}$, $g=0.59$, $f=f_\pi=132 \;
{\rm MeV}$ one
arrives at $\beta=2.3\pm 0.2 \; {\rm GeV}^{-1}$,
\footnote{The other solution is $\beta= -3.6 \pm 0.2 \; {\rm GeV}^{-1}$
that does not agree with $D^{*+}$ data.}
 where errors quoted  again reflect experimental errors only.
 The $\beta$ coupling coming from
from $D^{*+}$ \eqref{eq-102} and $D^{*0}$ \eqref{eq-108} are in
fair agreement, but  not equal. This signals that other
contributions
coming from   chiral loops and higher order terms that would alter our
determination of $\beta$ might be important. Since the contribution of
chiral
loops to $\Gamma(D^{*+}\to D^+ \gamma)$ are approximately $50\%$, while
for
$D^{*0}\to
D^0\gamma$ are about $20\%$ \cite{stewart}, we use
in our numerical calculations the value of
$\beta = 2.3\; {\rm GeV}^{-1}$
obtained from $\Gamma(D^{*0}\to D^0\gamma)$.
 Results are shown in Table \ref{tab-2}
using
$\beta=2.3 \; {\rm GeV}^{-1}$ and $m_c=1.4 \; {\rm GeV}$.
Inspection of Tables \ref{tab-1} and \ref{tab-2} reveals that for the
real
parts of the
amplitudes, the
contributions of Figs. \ref{fig-2}, \ref{fig-3} and of Fig. \ref{fig-5}
are
comparable in size. However, the
decay rate is dominated
by the contribution of the imaginary part of the parity-violating
amplitude, which arises from the
one loop diagrams of Fig. \ref{fig-2}. For the parity-conserving
amplitude, the
contributions of SD, anomaly
and $\beta$-like terms are comparable in magnitude.

Due to the suppression of $a_2$ in comparison to $a_1$, we do not
include diagrams proportional to $a_2$ in the calculation of terms with
$\beta$.

\begin{table} [h]
\begin{center}
\begin{tabular}{|l|c|c|c|} \hline
Diag. &$M_{i}^{(-)} [\times 10 ^{-10}{\rm  \;GeV}]$& $M_{i}^{(+)}
[\times 10
^{-10}{\rm  \;GeV^{-1}}]$\\ \hline\hline
$\beta .1$ & $0$ & $-2.69$ \\ \hline
$\beta .2$ & $0$ & $2.69$\\ \hline
$\beta .3$ &$0$ &  $2.11$ \\ \hline
$\beta .4$ & $0.88$ &$-0.007$ \\ \hline
$\beta .5$ &$0$ &  $0.51$ \\ \hline
$\beta .6$ &$-2.88 $&$-0.52$ \\ \hline
\hline
$\sum_i  M_i^{(\pm)}$& $-2.00$ &$2.09$\\ \hline

\end{tabular}
\caption{\footnotesize{Table of  nonzero contributions of the
amplitudes coming from the diagrams
 with $\beta$ coupling (Fig. \ref{fig-5}).
In the last line the sums of the contributions are  presented.
We use $\beta=2.3 $ GeV${}^{-1}$, $m_c=1.4$ GeV. }}
\label{tab-2}
\end{center}
\end{table}

Using short distance contributions, the finite one loop diagrams and the
anomaly parts of the
amplitudes (shown in Figs. \ref{fig-2}, \ref{fig-3}
and with numerical values of the amplitudes as listed in Table
\ref{tab-1}),
one obtains
\begin{equation}
Br(D^0\to \gamma\gamma)=1.0 \times 10^{-8}.
\end{equation}
This result is slightly changed when one takes into account the terms
dependent on $\beta$ \eqref{eq-100}. The branching ratio obtained when
we
sum all contributions is
\begin{equation}
 Br(D^0\to \gamma\gamma)=0.95 \times 10^{-8}.
\end{equation}

By varying $\beta$ within $1 \;{\rm GeV}^{-1}
\le \beta \le 5 \; {\rm GeV}^{-1}$ and keeping
$g=0.59\pm 0.08$, the branching ratio
is changed by at most 10\%. On the other hand, one has to keep in mind
that
the loop contributions
involving beta are not finite and have to be regulated. We have used
$\overline{\rm MS}$ scheme, with  the
 divergent parts being absorbed by counterterms.
 The size of these is not
known, so they might influence the error in our
 prediction of the branching ratio.
Note also  that changing $\alpha$ would
affect the predicted
branching ratio.  For instance, if the chiral corrections do
decrease the value of $\alpha$  by $30\%$ this would decrease
 the predicted
branching ratio down to $0.5\times 10^{-8}$.

\section{Summary}
\label{Summary}
We have presented here a detailed calculation of the decay amplitude
$D^0\to \gamma \gamma$, which
includes short distance and long distance contributions, by making use
of
the theoretical
tool of Heavy Quark Chiral Perturbation Theory Lagrangian. Within this
framework, the leading
contributions are found
to arise from the charged $\pi$ and $K$ mesons running in the chiral
loops, and
are  of order ${\cal O}(p^3)$.
These terms are finite and contribute only to the parity violating part
of
the amplitude. The
inclusion of terms of higher order in the chiral expansion is
unfortunately
 plagued
 with the uncertainty caused
by the lack of knowledge of the counterterms. As to the parity
conserving
part of the decay, it is given
by terms coming from the short distance contribution, the anomaly and
from
loop terms containing
the beta coupling, the latter giving most of the amplitude. The size of
this
part of the amplitude
is approximately one order of magnitude smaller than the parity
violating
amplitude, thus
contributing less than 20\% to the decay rate. Therefore, our
calculation
predicts that the $D\to 2 \gamma$
decay is mostly a parity violating transition.

 In addition to the uncertainties we have mentioned, there is the
question
of the suitability of
the chiral expansion for the energy involved in this process; the size
of
the uncertainty related
to this is difficult to estimate. Altogether, our estimate is that the
total
uncertainty is not larger
than 50\%. Accordingly, we conclude that the predicted branching ratio
is
\begin{equation}
Br(D^0\to \gamma \gamma)= (1.0 \pm 0.5)\times 10^{-8}. \label{fin-res}
\end{equation}
 The reasonability of this result can be deduced also from a comparison
with the calculated decay
rates for the $D^0\to \rho(\omega)\gamma$, which are found to be
expected
 with
a branching ratio of
approximately $10^{-6}$ \cite{FPS,BGHP,Fajfer-00}.

 We look forward to experimental attempts of detecting this decay. Our
result suggests that the
observation of $D\to 2 \gamma$ at a rate which is an order of magnitude
larger
than \eqref{fin-res}, could be a
signal for the type of "new physics", which leads to
sizable enhancement \cite{Sasathesis}
of the short-distance $c \to u \gamma$ transition.

\section*{Acknowledgments}
The research of S.F. and J.Z. was supported in part by the Ministry of
Education, Science and Sport of the Republic of Slovenia. The research
of P.S. was supported in part by Fund for Promotion of Research at the
Technion. P.S. also
acknowledges a helpful
communication from Prof. Ignacio Bediaga on the  $D^0\to 2 \gamma$
decay.

\appendix
\section{List of chiral loop integrals}
\label{app-A}
Here we list dimensionally regularized integrals needed in  evaluation
of
$\chi$PT and HQ$\chi$PT one-loop graphs shown on Fig.
\ref{fig-5}:
\begin{align}
i\mu^\epsilon\int \frac{d^{4-\epsilon}q}{(2\pi)^{4-\epsilon}
}\frac{1}{q^2-m^2}&=\frac{1}{16 \pi^2}I_1(m),\\
 i\mu^\epsilon\int \frac{d^{4-\epsilon}q}{(2\pi)^{4-\epsilon}}
\frac{1}{(q^2-m^2)(q\negcdot v-\Delta)}&=\frac{1}{16
\pi^2}\frac{1}{\Delta}I_2(m,\Delta), \label{last}
\end{align}
with
\begin{align}
I_1(m)&=m^2 \ln\Bigl(\frac{m^2}{\mu^2}\Bigr)-m^2\bar{\Delta} ,\\
I_2(m,\Delta)&=-2\Delta^2 \ln\Bigl(\frac{m^2}{\mu^2}\Bigr)-4 \Delta^2
F\Bigl(\frac{m}{\Delta}\Bigr) +2 \Delta^2(1+\bar{\Delta}),
\end{align}
where $\bar{\Delta}=\frac{2}{\epsilon}-\gamma +\ln(4\pi)+1$ (in
calculation $\bar{\Delta}=1$), while $F(x)$ is the function calculated
by Stewart in \cite{stewart}, valid for negative and positive values of
the argument
\begin{equation}
F\left(\frac{1}{x}\right)= \left\{
\begin{aligned}
-\frac{\sqrt{1-x^2}}{x}&\left[\frac{\pi}{2}-\tan^{-1}\left(\frac{x}{\sqrt{1-x^2}}\right)\right]\qquad
&|x|\le 1\\
\frac{\sqrt{x^2-1}}{x}&\ln \left(x+\sqrt{x^2-1}\right)\qquad &|x|\ge 1
\; .
\end{aligned}\right.
\end{equation}
The other integrals needed are (for $k^2=0$)
\begin{equation}
\begin{split}
i \int \frac{d^{4}q}{(2\pi)^{4}}&
\frac{1}{(q^2-m^2)((g+k)^2-m^2)(q\negcdot v-\Delta)}=\frac{1}{16
\pi^2}\frac{1}{v\negcdot k}G_0(m,\Delta,v\negcdot k),\\
& G_0(m,\Delta,v \negcdot k)=h^2(m,\Delta) -h^2(m,\Delta+v \negcdot
k)-\pi
[h(m,\Delta)-h(m,\Delta+v \negcdot k)],
\end{split}
\end{equation}
where
\begin{equation}
 h(m,\Delta)=\left\{
\begin{aligned}
\> &\Arctan \left(\frac{\Delta}{\sqrt{m^2-\Delta^2}}\right) ; |m|>
|\Delta|\\
i & \ln \left|
\frac{m}{\Delta-\sqrt{\Delta^2-m^2}}\right|+\sign(\Delta)\frac{\pi}{2};
|m|<|\Delta|
\end{aligned}\right. ,
\end{equation}
and
\begin{equation}
\begin{split}
i\mu^\epsilon \int \frac{d^{4-\epsilon}q}{(2\pi)^{4-\epsilon}}&
\frac{q^\mu q^\nu}{(q^2-m^2)((g+k)^2-m^2)(q\negcdot
v-\Delta)}=\frac{1}{16\pi^2}\Big[ g^{\mu \nu}G_3(m,\Delta,v\negcdot
k)+\\
&(v^\mu k^\nu +k^\mu v^\nu) G_4(m,\Delta,v \negcdot k)+k^\mu k^\nu G_5
(m,
\Delta, v\negcdot k) +v ^\mu v^\nu G_6 (m, \Delta, v \negcdot k)\Big],
\end{split}
\end{equation}
with
\begin{subequations}
\begin{align}
G_3(m,\Delta, v\negcdot k)&= \frac{m^2}{2 v \negcdot k} G_0 (m,\Delta,
v\negcdot k) - \frac{1}{4 v\negcdot k}\Big[ I_2(m,\Delta)-I_2
(m,\Delta+v
\negcdot k)\Big] +\Delta + \frac{v \negcdot k}{2},\\
G_4(m,\Delta, v \negcdot k)&= \frac{1}{v \negcdot k} \Big[
\frac{1}{2(\Delta +
v \negcdot k)} I_2 (m, \Delta +v \negcdot k) - G_3 (m, \Delta , v
\negcdot
k)\Big],\\
G_5(m,\Delta, v \negcdot k )&= \frac{1}{v \negcdot k} \Big[
-\frac{1}{2}N_0(m,0) +\Delta G_2 (m,\Delta, v \negcdot k)- G_4(m, \Delta
+v
\negcdot k)\Big],\\
G_6(m,\Delta, v\negcdot k)&=\frac{1}{2 v \negcdot k}\Big[ I_2(m,\Delta)-
I_2(m,\Delta+v \negcdot k)\Big],
\end{align}
\end{subequations}
where $I_2(m,\Delta)$ is defined in \eqref{last} and $N_0(m,k^2)$ in
\eqref{eq-111}.

In calculation we also need several other integrals that have been
calculated
for the case $k_1^\mu+k_2^\mu=m_D v^\mu$, $k_1^2=k_2^2=0$, $v\negcdot
k_1=
v\negcdot k_2=\tfrac{m_D}{2}$
\begin{align}
i \int \frac{d^{4}q}{(2\pi)^{4}}&
\frac{1}{((q+k_1)^2-m^2)((q-k_2)^2-m^2)(q\negcdot v-\Delta)}=\frac{1}{16
\pi^2}\;\overline{G}_0(m,\Delta,m_D),\\
i \int \frac{d^{4}q}{(2\pi)^{4}}&
\frac{1}{((q+k_1)^2-m^2)((q-k_2)^2-m^2)(q^2-m^2)(q\negcdot
v-\Delta)}=\frac{1}{16
\pi^2}\;\overline{M}_0(m,\Delta,m_D),
\end{align}
where
\begin{align}
\begin{split}
\bar{G}_0(m,\Delta,m_D)=\frac{2 }{\Delta m_D} \Big[ &\Big(\tfrac{\pi}{2}
-
h(m,\Delta-\tfrac{m_D}{2})\Big)\sqrt{m^2 -
(\Delta-\tfrac{m_D}{2})^2- i \delta \;} \\
&-\Big(\tfrac{\pi}{2} -
h(m,\Delta+\tfrac{m_D}{2})\Big)\sqrt{m^2 - (\Delta+\tfrac{m_D}{2})^2- i
\delta \;} \\
&-2 h(m,\tfrac{m_D}{2}) \sqrt{m^2- \tfrac{m_D^2}{4} - i \delta \;}\;
\Big],
\end{split}\\
\begin{split}
\bar{M}_0(m,\Delta,m_D)=\frac{1}{\Delta m_D^2} \Bigg\{& -2
h^2(m,\tfrac{m_D}{2}) -2 h^2(m,\Delta) + h^2(m,\Delta-\tfrac{m_D}{2})
+h^2(m,\Delta+\tfrac{m_D}{2})\\
&+i \pi \ln \left[\frac{\Delta-\tfrac{m_D}{2} - i \sqrt{m^2-
(\Delta-\tfrac{m_D}{2})^2-i \delta \;}}{-\Delta-\tfrac{m_D}{2} - i
\sqrt{m^2-
(\Delta+\tfrac{m_D}{2})^2-i \delta \;}}\right]+\\
&+ i \pi \ln \left[\frac{\Delta+ i \sqrt{m^2- \Delta^2-i
\delta \;}}{-\Delta + i
\sqrt{m^2-\Delta^2-i \delta \;}}\right]\Bigg\},
\end{split}
\end{align}
with $\delta>0$ an infinitesimal positive parameter.\\
The chiral loops needed are
\begin{equation}
\begin{split}
i\mu^\epsilon\int &\frac{d^{4-\epsilon}q}{(2\pi)^{4-\epsilon}}
\frac{1}{(q^2-m^2)((q+k)^2-m^2)}=\frac{1}{16\pi^2}N_0(m,k^2),\\
& N_0(m,k^2)=-\bar{\Delta} +1 - H\big(\tfrac{k^2}{m^2}\big)
+\ln\big|\tfrac{m^2}{\mu^2}\big|,\label{eq-111}
\end{split}
\end{equation}
where
\begin{equation}
H(a)= \left\{
\begin{aligned}
2&\Bigg(1-\sqrt{4/a-1}\;
\Arctan\bigg(\frac{1}{\sqrt{4/a-1}}\bigg)\Bigg)\quad ; 0<a<4\\
2&\Bigg(1-\sqrt{1-4/a}\;\frac{1}{2}\Bigg\{\ln\Bigg|\frac{\sqrt{1-4/a}+1}{\sqrt{1-4/a}-1}\Bigg|-i\pi
\Theta(a-4)\Bigg\}\Bigg)\quad ; {\rm otherwise}
\end{aligned}\right. ,
\end{equation}
and for $k_1^2=k_2^2=0$
\begin{equation}
\begin{split}
i\mu^\epsilon\int &\frac{d^{4-\epsilon}q}{(2\pi)^{4-\epsilon}}
\frac{q^\mu
q^\nu}{(q^2-m^2)((q+k_1)^2-m^2)((q+k_2)^2-m^2)}=-\frac{1}{16\pi^2}\bigg[g^{\mu
\nu}M_2(m,k_1\negcdot k_2)\\
-&\frac{k_1^\mu k_1^\nu+k_2^\mu k_2^\nu}{k_1\negcdot k_2}
M_3(m,k_1\negcdot
k_2) -\frac{k_1^\mu k_2^\nu +k_1^\nu k_2^\mu}{k_1\negcdot
k_2}M_4(m,k_1\negcdot
k_2)\bigg],
\end{split}
\end{equation}
with
\begin{subequations}
\begin{align}
\begin{split}
M_2(m, k_1\negcdot k_2)=\frac{1}{2}\bigg[& \frac{1}{2}
\left(\bar{\Delta}-\ln\left(\frac{m^2}{\mu^2}\right)\right)+\frac{1}{a}\bigg(\Li_2\Big(\frac{2}{1+\sqrt{\quad}}\Big)+\Li_2\Big(\frac{2}{1-\sqrt{\quad}}\Big)\\
&+1-\sqrt{\quad}\Arcth\bigg(\frac{1}{\sqrt{\quad}}\bigg)\bigg],
\end{split}\\
M_3(m,k_1\negcdot
k_2)=\frac{1}{2}\bigg[&\sqrt{\quad}\Arcth\Big(\frac{1}{\sqrt{\quad}}\Big)-1\bigg),\\
M_4(m,k_1\negcdot
k_2)=\frac{1}{4}+&\frac{1}{2a}\bigg[\Li_2\Big(\frac{2}{1+\sqrt{\quad}}\Big)+\Li_2\Big(\frac{2}{1-\sqrt{\quad}}\Big)\bigg],
\end{align}
\end{subequations}
where we have abbreviated $a=2 k_1\negcdot k_2/m^2$ and
$\sqrt{\quad}=\sqrt{1+2
m^2/k_1\negcdot k_2}$, while $\Li_2(x)$ is a polylogarithmic function.


\begin{thebibliography}{99}
\bibitem{CLEO1} R. Godang et al. (CLEO Collaboration),
Phys. Rev. Lett. {\bf 84},  5038 (2000).
\bibitem{FOCUS1} J.~M. Link et al. (FOCUS Collaboration),
Phys. Lett. B {\bf 485},  62 (2000).
\bibitem{Nir} S.~Bergmann, Y.~Grossman, Z.~Ligeti, Y.~Nir and
A.~A.~Petrov,
Phys.\ Lett.\ B {\bf 486}, 418 (2000);
I. I. Bigi and N. G. Uraltsev, Nucl. Phys.
B {\bf 592}, 92 (2001).
\bibitem{CLEO2}T. E. Coan et al.
(CLEO Collaboration), hep-ex/0102007, M. Dubrovin
(for CLEO Collaboration),
hep-ex/0105030.
\bibitem{E791} A.~Freyberger et al. (CLEO
Collaboration), Phys. Rev. Lett. {\bf 76}, 3065 (1996); D.~M.~Asner et
al.  (CLEO Collaboration),
Phys. Rev. D {\bf 58}, 092001 (1998); E.~M.~Aitala et al. (E791
Collaboration), hep-ex/0011077; D.~A.~Sanders,
Mod. Phys. Lett. A {\bf 15}, 1399 (2000); A.~J.~Schwartz,
hep-ex/0101050; D.~J.~Summers, hep-ex/0011079.
\bibitem{FPS} S. Fajfer, S. Prelov\v sek, P. Singer,
Eur. Phys. J. C {\bf 6}, 471, 751(E) (1999).
\bibitem{BGHP} G. Burdman, E. Golowich, J. L. Hewett and S. Pakvasa,
Phys. Rev. D {\bf 52}, 6383 (1995).
\bibitem{HP} Q. Ho-Kim , X.~Y. Pham,
Phys. Rev. D {\bf 61},  013008 (2000).
\bibitem{Khodj} A.~Khodjamirian, G.~Stoll and D.~Wyler, Phys. Lett. B
{\bf 358},129 (1995).
\bibitem{Fajfer-98} S.~Fajfer, S.~Prelov\v sek and P.~Singer, Phys. Rev.
D {\bf 58}, 094038 (1998).
\bibitem{Lebed-00} R.~F.~Lebed, Phys. Rev. D {\bf 61}, 033004 (2000).
\bibitem{Geng} C. Q. Geng, C. C. Lih, W-M. Zhang,
Mod. Phys. Lett. A {\bf 15}, 2087 (2000).
\bibitem{Fajfer-00} S.~Fajfer, S.~Prelov\v sek, P.~Singer and D.~Wyler,
Phys. Lett. B {\bf  487}, 81 (2000).
\bibitem{RG} H. Routh, V. P. Gautam, Phys. Rev. D {\bf 54}, 1218 (1996);
H.~Routh, H.~Roy, A.~K.~Maity
and V.~P.~Gautam,
Acta Phys. Pol. B {\bf 30}, 2687 (1999).
\bibitem{LSH} G.-L. Lin, J. Liu, Y.-P. Yao, Phys. Rev. Lett.
{\bf 64}, 1498 (1990); {\it ibid.}, Phys. Rev. D {\bf 42}, 2314 (1990);
{\it ibid.}, Mod. Phys. Lett. A {\bf 6},  1333 (1991);
H. Simma, D. Wyler, Nucl. Phys. B {\bf 344}, 283 (1990);
E. Vanem and J. O. Eeg, Phys. Rev. D {\bf 58}, 114010 (1998).
\bibitem{RRS} L. Reina, G. Riccardi, A. Soni,
Phys. Rev. D {\bf 56}, 5805 (1997); G.~Hiller and
E.~O.~Iltan, Phys. Lett. B {\bf 409},
425 (1997).
\bibitem{BI} M. Boz, E. O. Iltan, Phys. Rev. D {\bf 62},
054010 (2000).
\bibitem{ELLIS} D. Choudhury, J. Ellis, Phys. Lett. B {\bf 433},
 102  (1998); W. Liu, B. Zhang and H. Zheng, Phys. Lett.
 B {\bf 461},  295 (1999).
\bibitem{Eilam} G.~Eilam, A.~Ioannissian, R.~R.~Mendel and
P.~Singer, Phys. Rev. D {\bf 53}, 3629 (1996); A.~Ali, DESY-97-192,
hep-ph/9709507.
\bibitem{Gaillard} M.~K.~Gaillard and B.~W.~Lee, Phys. Rev. D {\bf 10},
897 (1974); E.~Ma and
A.~Pramudita, Phys. Rev. D {\bf 24}, 2476 (1981);
J.~O.~Eeg, B.~Nizic and I.~Picek, Phys. Lett. B {\bf 244}, 513 (1990).
\bibitem{Goity} J. L. Goity, Z. Phys. C {\bf 34}, 341  (1987).
\bibitem{Amb} G. D'Ambrosio and D. Espiriu, Phys. Lett. B {\bf 175},
 237 (1986).
\bibitem{Kamb} J. Kambor and B. R. Holstein, Phys. Rev. D {\bf 49},
 2346 (1994).
\bibitem{wise}
 M.\ B.\ Wise, Phys.\ Rev.\ D {\bf 45}, 2188 (1992); G.~Burdman and
J.~Donoghue, Phys. Lett. B {\bf 280}, 287 (1992).
\bibitem{itchpt}
R.\ Casalbuoni,
A.\ Deandrea, N.\ Di Bartolomeo, R.\ Gatto, F.\ Feruglio,
 and  G.\ Nardulli, Phys.\ Rep.\ {\bf 281}, 145 (1997). This review
provides
many references on the use of HQCT
in different processes.
\bibitem{stewart}
 I.\ W.\ Stewart, Nucl.\ Phys.\  B  {\bf 529}, 62 (1998).
\bibitem{GS} D. Guetta, P. Singer, Phys. Rev. D {\bf 61},
 054014  (2000).
\bibitem{HK} S. Herrlich, J. Kalinowski,  Nucl. Phys. B {\bf 381},
502 (1992).
\bibitem{GMW} C. Greub, T. Hurth, M. Misiak and D. Wyler,
Phys. Lett. B {\bf 382},  415 (1996).
\bibitem{Sasathesis} S. Prelov\v sek, D. Wyler, Phys. Lett. B {\bf 500},
304 (2001); S.~Prelov\v sek,
hep-ph/0010106.
\bibitem{buras}
 A.\ J.\ Buras, Nucl.\ Phys.\ B {\bf 434}, 606 (1995).
\bibitem{Eeg:2001un}
J.~O.~Eeg, S.~Fajfer and J.~Zupan,
hep-ph/0101215, to appear in Phy. Rev. D.
\bibitem{CLEO}
G.\ Bonvicini et al. (CLEO Collaboration), Phys. Rev. D {\bf  63},
071101 (2001).
\bibitem{PDG-00}
 Review of Particle Physics, D. E.\ Groom et al., Eur.
Phys.\ J. C {\bf 15}, 1 (2000).
\bibitem{Grinstein-92}
 B.\ Grinstein, E. Jenkins, A. V. Manohar, M. J. Savage, M. B. Wise,
Nucl.\ Phys.\  B {\bf  380}, 369 (1992).
\bibitem{Grinstein-94}
 A.\ F.\ Falk,  B.\ Grinstein, Nucl.\ Phys.\ B {\bf 416}, 771 (1994).
\bibitem{BG} C.\ G.\ Boyd and B.\ Grinstein, Nucl.\ Phys. B
{\bf 442}, 205 (1995).





\end{thebibliography}
\end{document}